\documentclass[10pt, final, journal, letterpaper,twocolumn]{IEEEtran}
\usepackage{graphicx}
\usepackage{subfigure,color}
\usepackage{amssymb}
\usepackage{cite,comment}
\usepackage{amsmath}
\usepackage{bm,comment}
\usepackage{algorithm}
\usepackage{algpseudocode}

\usepackage{stfloats}

\newtheorem{remark}{Remark}

\begin{document}
\title{A New Store-then-Amplify-and-Forward Protocol for UAV Mobile Relaying}
\author{
Xiaochen Lin, Weidong Mei, and Rui Zhang, \IEEEmembership{Fellow, IEEE}
\thanks{This work was supported by the Research and Innovation Project of Shanghai Municipal Education Commission (No.15ZZ105). X. Lin is with School of Electronic Information Engineering, Shanghai Dianji University, Shanghai 201306, China (e-mail: linxc@sdju.edu.cn).
W. Mei and R. Zhang are with the Department of Electrical and Computer Engineering, National University of Singapore (NUS), Singapore 117583 (e-mail:wmei@u.nus.edu, elezhang@nus.edu.sg). W. Mei is also with the NUS Graduate School for Integrative Sciences and Engineering, NUS, Singapore 119077.
}
}
\maketitle

\begin{abstract}
In this letter, we consider the use of an unmanned aerial vehicle (UAV) as a mobile relay to assist the communication between two ground users without a direct link. We propose a novel store-then-amplify-and-forward (SAF) relaying protocol for the UAV to exploit its mobility jointly with the low-complexity AF relaying. Specifically, the received signal from the source is first stored in a buffer at the UAV, then amplified and forwarded to the destination when the UAV flies closer to the destination. With this new SAF protocol, we aim to maximize the throughput of the UAV-enabled relaying system by jointly optimizing the source/UAV transmit power and the UAV trajectory, as well as the time-slot pairing for each data packet received and forwarded by the UAV. As this problem is a non-convex mixed integer optimization problem that is difficult to solve, we propose an efficient algorithm for obtaining a suboptimal solution for it by applying the techniques of Hungary algorithm, alternating optimization and successive convex approximation. Numerical results show that the proposed mobile SAF relaying outperforms the conventional AF relaying without signal storing.
\end{abstract}


\vspace{-0.5cm}
\section{Introduction}
Thanks to their high mobility, flexible deployment and line-of-sight (LoS)-dominant links with ground nodes, unmanned aerial vehicles (UAVs) are anticipated to be widely used for aiding the future wireless communication systems \cite{ZengandZhangComMag16}. In particular, mounted with miniaturized BSs/relays, UAVs can be deployed as aerial mobile communication platforms to provide or enhance the communication services for terrestrial users. Furthermore, the high mobility of UAVs also offers a new degree of freedom for communication performance improvement, via optimizing their locations over time, termed trajectory design. By proper UAV trajectory design, the communication distances between the UAV and its served ground users can be greatly  shortened, thus improving their communication performance significantly.

One UAV application of high practical interest is UAV-enabled mobile relaying, where UAVs are deployed to assist the communication between terrestrial nodes, which has been investigated in the literature \cite{ZengandZhangTCom16,ZhangAccess18,AlsharoaICC18,ZhangandZhangTSP18,ZhangComletter17,JiangAccess18} under different relaying protocols such as decode-and-forward (DF) \cite{ZengandZhangTCom16,ZhangAccess18,AlsharoaICC18,ZhangandZhangTSP18} and amplify-and-forward (AF) \cite{ZhangComletter17,JiangAccess18}. The work \cite{ZengandZhangTCom16} showed that UAV-enabled mobile relaying with proper trajectory design is able to significantly improve the communication throughput as compared to the conventional DF relay at a fixed location, by letting the UAV relay transmit/receive when it flies closer to the destination and the source, respectively. Furthermore, in \cite{ZhangAccess18} and \cite{AlsharoaICC18}, DF-based UAV relaying was extended to the more general case of multiple UAV relays; while in \cite{ZhangandZhangTSP18}, the case with multiple source-destination pairs was considered. In contrast to DF relaying, the works in \cite{ZhangComletter17} and \cite{JiangAccess18} studied UAV-enabled AF relaying, for minimizing the communication outage probability and maximizing the communication throughput, respectively, by jointly optimizing the UAV transmit power and trajectory. However, these two works considered that the UAV forwards its received signal immediately without data buffering, and as a result, did not fully exploit the UAV mobility or trajectory design for performance optimization. Instead, with signal buffering at the UAV, its relaying performance can be further improved, e.g., when it is far from the destination, the received signal from the source can be temporarily stored in a buffer, and sent to the destination when the UAV flies closer to it and thus has a better channel condition with it.

Motivated by the above, this letter proposes a new {\it store-then-amplify-and-forward} (SAF) relaying protocol, where the UAV can store its received signal from the source for a certain duration before forwarding it to the destination. Assuming that the UAV relay operates in frequency division duplexing (FDD) mode, the new SAF relaying protocol needs to determine an additional set of design variables for pairing the time-slots for each data packet received/transmitted by the UAV, as compared to the conventional AF protocol with instant relaying (thus termed instant AF (IAF) in the sequel). Under the proposed new SAF relaying protocol, we aim to maximize the system throughput by jointly optimizing the UAV receive/transmit time-slot pairing, the source/UAV transmit power and the UAV trajectory, subject to various practical constraints. Due to the new constraints on time-slot pairing, this throughput maximization problem is a mixed integer non-convex optimization problem that is challenging to solve. To tackle this challenge, we propose an efficient iterative algorithm by applying the technique of alternating optimization (AO). It is shown by simulation results that the proposed mobile SAF relaying outperforms the conventional IAF relaying.

\vspace{-0.3cm}
\section{System Model and Problem Formulation}
\subsection{System Model}
\begin{figure}[t]
\centering
\includegraphics[width=1.8in]{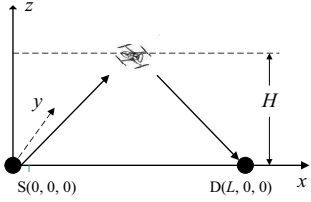}  \vspace{-0.3cm}
\caption{A UAV-enabled wireless relaying system.} \vspace{-0.5cm}
\label{fig:model}
\end{figure}
As shown in Fig. \ref{fig:model}, we consider a mobile relaying system consisting of a pair of source (S) and destination (D) nodes at fixed locations on the ground and a UAV-mounted relay. The direct link between S and D is assumed to be negligible due to the long distance or severe blockage between them. Without loss of generality, we consider a three-dimensional (3D) Cartesian coordinate system with S and D located at $(0,0,0)$ and $(L,0,0)$ in meter (m), respectively. Moreover, we assume that the UAV flies at a fixed altitude $H$ during a flight period $T$ and has a time-varying coordinate denoted by $(x(t),y(t),H), 0 \leq t \leq T$. In practice, the flight period $T$ can be determined based on the UAV's maximum endurance. For simplicity, we divide $T$ into a sufficiently large number of equal time-slots denoted by $N$, such that the UAV trajectory can be well approximated by a sequence of line segments connecting the way points denoted by $(x[i],y[i],H), i \in {\cal N} \triangleq \{1,2,\cdots,N\}$. Moreover, the UAV has pre-determined lauching/landing locations, denoted as $(x_0,y_0,H)$ and $(x_F,y_F,H)$, respectively, while they can be ignored if free initial and final locations are considered. Furthermore, we define $D_u \triangleq V_u(T/N)$ as the UAV's maximum displacement within a single time-slot, where $V_u$ denotes the UAV maximum speed. Consequently, the UAV trajectory $\{x[i],y[i]\}^{N}_{i=1}$ should satisfy the following constraints:
\begin{align}
(x[1]-x_0)^2+(y[1]-y_0)^2&\leq D_u^2,\label{maxV1}\\
(x[i+1]\!-\!x[i])^2+(y[i+1]\!-\!y[i])^2\!&\leq D_u^2, i \in {\cal N} \backslash \{N\}, \label{maxV2} \\
(x_F-x[N])^2+(y_F-y[N])^2&\leq D_u^2.\label{maxV3}
\end{align}

Denote by $h_{su}[i]$ and $h_{ud}[i]$ the channel power from S to the UAV, and that from the UAV to D in the $i^{\text{th}}$ time-slot, respectively. Thus, in the $i^{\text{th}}$ time-slot, the received signal at the UAV can be expressed as
\begin{align}
y_{u}[i] = \sqrt{P_{s}[i]h_{su}[i]}x_{s}[i] + n_{u}[i],
\end{align}
where $x_{s}[i], P_{s}[i]$ and $n_{u}[i]$ denote the transmitted symbol from S, the transmit power of S and the receiver Gaussian noise with power $\sigma^2$ in the $i^{\text{th}}$ time-slot, respectively. Due to the LoS-dominant channels and for the ease of the offline trajectory design, we assume for simplicity the free-space path-loss model in this work as in \cite{ZengandZhangTCom16} and \cite{ZhangAccess18}, i.e.,
$h_{su}[i]=\beta_{0}d^{-2}_{su}[i]=\frac{\beta_0}{x^2[i]+y^2[i]+H^2}$, where $\beta_0$ denotes the channel power at the reference distance of 1 m and $d_{su}[i]$ is the distance between S and the UAV.

This letter considers that the UAV applies AF relaying with lower implementation complexity but higher data privacy than DF. For the proposed SAF scheme, we assume that the data buffer at the UAV is sufficiently large to store its received signals from S before sending them to D. Specifically, with FDD, if the UAV assigns the $j^{\text{th}}$ time-slot to forward the signal received in the $i^{\text{th}}$ time-slot with $j \ge i$, we refer to these two time-slots as a pair of SAF time-slots. Note that if $j=i$ holds for each paired $i^{\text{th}}$ and $j^{\text{th}}$ time-slots, $i, j \in \cal N$, the proposed SAF scheme reduces to the conventional IAF relaying scheme. As a result, SAF includes IAF as a special case and is anticipated to yield a better performance than IAF. To characterize different pairings for the $N$ time-slots, we define the following binary variables for all $i,j \in {\cal N}$, i.e.,

\vspace{-0.2cm}
\begin{align}
w[i][j]=
\begin{cases}
1, &\text{if the $i^{\text{th}}$ and $j^{\text{th}}$ time-slots are paired,} \\
0, &\text{otherwise.}
\end{cases}
\end{align}

\vspace{-0.1cm}
To ensure that the received signal from S at any $i^{\text{th}}$ time-slot can be forwarded to D in only one of the current and subsequent time-slots, the following constraints must be satisfied, i.e.,
\vspace{-0.2cm}
\begin{align}\label{jaddequal1}
\underset{j=i}{\overset{N}{\sum}} w[i][j] \leq 1,  ~~\forall i \in \cal N.
\end{align}
Moreover, in any given $j^{\text{th}}$ time-slot, the UAV can only forward the signal received from at most one of the current and previous time-slots, which yields the following constraints:
\vspace{-0.2cm}
\begin{align}\label{iaddequal1}
\underset{i=1}{\overset{j}{\sum}} w[i][j] \leq 1, ~~\forall j \in \cal N.
\end{align}

Based on the above, if $w[i][j]=1$, in the $j^{\text{th}}$ time-slot, the UAV will transmit an amplified version of the signal received in the $i^{\text{th}}$ time-slot , $i\leq j$. The signal transmitted by the UAV in the $j^{\text{th}}$ time-slot can be expressed as
\vspace{-0.1cm}
\begin{align}
x_{u}[j] &= \alpha[i][j]y_{u}[i]  \nonumber \\
         &= \alpha[i][j]\sqrt{P_{s}[i]h_{su}[i]}x_{s}[i] + \alpha[i][j]n_u[i],
\end{align}
where $\alpha[i][j] = \sqrt{\frac{P_{u}[j]}{P_{s}[i]h_{su}[i]+\sigma^2}}$ denotes the amplification coefficient and $P_{u}[j]$ is the UAV transmit power in the $j^{\text{th}}$ time-slot. Thus, the received signal at D is
\vspace{-0.1cm}
\begin{align}
y_{d}[j]&=\sqrt{h_{ud}[j]}x_{u}[j] \!+ \!n_{d}[j]  \\
        &=\alpha[i][j](\sqrt{h_{ud}[j]P_{s}[i]h_{su}[i]}x_{s}[i]\!+ \!\sqrt{h_{ud}[j]}n_{u}[i])\!+\! n_{d}[j],  \nonumber
\end{align}
where $n_d[j]$ denotes the receiver Gaussian noise at D in the $j^{\text{th}}$ time-slot and the UAV-to-D channel power $h_{ud}[j]$ is also assumed to follow the free-space path-loss model, i.e.,
$ h_{ud}[j] = \beta_{0}d^{-2}_{ud}[j] = \frac{\beta_0}{(x[j]-L)^2+y^2[j]+H^2}$, with $d_{ud}[j]$ denoting the distance between D and the UAV.

As a result, the signal-to-noise ratio (SNR) at D in the $j^{\text{th}}$ time-slot can be expressed as
\vspace{-0.1cm}
\begin{align}
{\text{SNR}}[i][j] &= \frac{ \alpha^2[i][j]h_{ud}[j]P_{s}[i]h_{su}[i] }{(\alpha^2[i][j]h_{ud}[j]+1)\sigma^2 } \nonumber \\
          &= \frac{ P_{u}[j]h_{ud}[j]P_{s}[i]h_{su}[i] }{ (P_{u}[j]h_{ud}[j] + P_{s}[i]h_{su}[i])\sigma^2 + \sigma^4}.
\end{align}
Accordingly, the achievable rate in bits/second/Hertz (bps/Hz) for the paired $i^{\text{th}}$ and $j^{\text{th}}$ time-slots can be expressed as
\vspace{-0.1cm}
\begin{align}
R[i][j]=\log_2(1+{\text{SNR}}[i][j]).
\end{align}

\subsection{Problem Formulation}
In this letter, we aim to maximize the throughput of the considered relaying system under the SAF protocol, by jointly optimizing the time-slot pairing, the source/UAV transmit power over time and the UAV trajectory. Thus, the optimization problem is formulated as
\vspace{-0.1cm}
\begin{align}
\mathrm{(P1):} \underset{\{x[i],y[i]\}}{ {\underset{\{w[i][j],P_{s}[i],P_{u}[j]\}}{\max}}}
& ~ \underset{i=1}{\overset{N}{\sum}} \underset{j=i}{\overset{N}{\sum}} ~ w[i][j]R[i][j]  \nonumber\\
\mathrm{s.t.}  ~~&w[i][j] \in \{0,1\}, ~ \forall ~i,j \in {\cal N},  \label{w01} \\
&  w[i][j]=0, ~ \forall ~i,j \in {\cal N}, ~j<i, \label{w=0}\\
&  \underset{i=1}{\overset{N}{\sum}} P_{s}[i] \leq E_{s}, \underset{j=1}{\overset{N}{\sum}} P_{u}[j] \leq E_{u},  \label{average_power}\\
&  P_{s}[i] \geq 0,  ~P_{u}[i] \geq 0, ~ \forall i \in {\cal N}, \label{power_morethan0} \\
& (\ref{maxV1}), (\ref{maxV2}), (\ref{maxV3}) ,(\ref{jaddequal1}) , (\ref{iaddequal1}), \nonumber
\end{align}
where $E_s$ and $E_u$ denote the total energy budget of S and the UAV during the flight, respectively. Note that (P1) is a mixed integer non-convex optimization problem and thus it is difficult to solve (P1) optimally in general.

\begin{remark}
To characterize the optimal performance of the proposed SAF scheme, we do not constrain the UAV's buffering delay in (P1). Nonetheless, our problem formulation is also applicable to the scenarios with additional constraints on the buffering delay. Specifically, let $D_m \ge 0$ be the maximum allowable buffering delay (in time slots) in the considered system. To meet this new requirement, we only need to add the following constraints in (P1), i.e. $w[i][j]=0,~ \forall ~i,j \in {\cal N},~j>i+D_m$. As these constraints are equality constraints, the resulting problem can still be solved via the proposed algorithm to be presented in the next section.
\end{remark}
\vspace{-0.3cm}

\section{Proposed Algorithm}
In this section, we propose an efficient iterative algorithm by decomposing (P1) into three sub-problems. The main idea is to iteratively maximize the throughput by optimizing the time-slot pairing, the source/UAV transmit power and the UAV trajectory in an alternating manner.

\vspace{-0.3cm}

\subsection{Time-Slot Pairing Optimization}
For any given source/UAV transmit power and UAV trajectory, (P1) can be reduced to the following sub-problem:
\begin{align}
\text{(P1.1)}: \underset{\{w[i][j]\}}{\max}~&
   \underset{i=1}{\overset{N}{\sum}} \underset{j=i}{\overset{N}{\sum}} ~ w[i][j]R[i][j] \nonumber\\
\mathrm{s.t.}~~&\text{(\ref{jaddequal1}),(\ref{iaddequal1}),(\ref{w01}),(\ref{w=0})}. \nonumber
\end{align}
Note that (P1.1) is a standard 0-1 integer linear programming. Such a problem is known equivalent to the classical bipartite matching problem in graph theory, and thus can be optimally solved by the Hungary algorithm\cite{Bertsekas}.

\vspace{-0.3cm}
\subsection{Power Allocation Optimization}\label{pw.opt}
For any given time-slot pairing and UAV trajectory, the sub-problem of transmit power optimization is given by
\begin{align}
\text{(P1.2)}:
\underset{\{P_{s}[i],P_{u}[j]\}}{\max}~&
\underset{i=1}{\overset{N}{\sum}} \underset{j=i}{\overset{N}{\sum}}
w[i][j]f_1(P_s[i],P_u[j]) \nonumber \\
\rm{s.t.} ~~&\;\text{(\ref{average_power}), (\ref{power_morethan0})}, \nonumber
\end{align}
where \begin{small}
$f_1(P_s[i],P_u[j])=\log_2\left(1+\frac{P_u[j]\rho_u[j]P_s[i]\rho_s[i]}{P_u[j]\rho_u[j]+P_s[i]\rho_s[i]+1}\right),$
\end{small} with \begin{small}$\rho_u[j]\!=\!\frac{\beta_0}{((x[j]-L)^2+y^2[j]+H^2)\sigma^2}$ \end{small} and \begin{small}$\rho_s[i]\!=\!\frac{\beta_0}{(x^2[i]+y^2[i]+H^2)\sigma^2}$\end{small}.
Note that $f_1(P_s[i],P_u[j])$ is non-concave with respect to (w.r.t.) $P_{s}[i]$ and $P_{u}[j]$. Thus, problem (P1.2) is a non-convex optimization problem.

To tackle this problem, we introduce the following variable transformation, i.e., $T_s[i]\triangleq 1/P_s[i]$ and $T_u[j]\triangleq 1/P_u[j]$ for all $(i,j)$'s satisfying $w[i][j]=1$. As a consequence, the objective function of (P1.2) can be recast as
\begin{small}
$f_2(T_s[i],T_u[j])
=\log_2\left(1+\frac{\rho_s[i] \rho_u[j]}{\rho_u[j]T_s[i]+\rho_s[i]T_u[j]+T_s[i]T_u[j]}\right).$
\end{small}
It can be verified that $f_2(T_s[i],T_u[j])$ is a jointly convex function w.r.t. $T_s[i]$ and $T_u[j]$, since its Hessian matrix is positive semi-definite (for which the proof is omitted due to the space limit). It is known that the first-order Taylor approximation of a convex function $f(a)$ is its global under-estimator, i.e.,\begin{small} $f(a)\geq f(a_0)+\nabla f(a_0)^T (a-a_0)$\end{small}, where $\nabla f(a_0)$ is the gradient of $f(a)$ at $a_0$.
Therefore, we can apply the successive convex approximation (SCA) technique to find a locally optimal solution to (P1.2). Specifically, define $T_{s,l}[i]$ and $T_{u,l}[j]$ as the given local points in the $l^{\text{th}}$ SCA iteration. Then $f_2(T_s[i],T_u[j])$ can be approximated by the following lower bound in the $l^{\text{th}}$ iteration,
\begin{align}
f_{2,\text{LB}}(T_s[i],T_u[j])\!\triangleq\!& f_2(T_{s,l}[i],\!T_{u,l}[j])\!-\!A_{s,l}[i][j](T_s[i]\!-\!T_{s,l}[i])\nonumber\\
&-A_{u,l}[i][j](T_u[j]-T_{u,l}[j]),
\end{align}
where
\begin{small}
\begin{align}
A_{s,l}[i][j] &= \frac{\rho_s[i]\rho_u[j]}{D[i][j](T_{s,l}[i]+\rho_s[i])\ln2}> 0, \\
A_{u,l}[i][j] &= \frac{\rho_s[i]\rho_u[j]}{D[i][j](T_{u,l}[j]+\rho_u[j])\ln2}> 0,
\end{align}
\end{small}
with \begin{small}$D[i][j] \triangleq \rho_u[j]T_{s,l}[i]+\rho_s[i]T_{u,l}[j]+T_{s,l}[i]T_{u,l}[j], i,j \in \cal N$\end{small}.

Accordingly, in the $l^{\text{th}}$ SCA iteration, we need to solve the following optimization problem
\begin{align}
\text{(P1.3)}: \underset{\{T_s[i],T_u[j]\}}{\max}~&\underset{i=1}{\overset{N}{\sum}} \underset{j=i}{\overset{N}{\sum}} ~w[i][j] ~f_{2,\text{LB}} \nonumber \\
\mathrm{s.t.}~&
\underset{i=1}{\overset{N}{\sum}} \frac{1}{T_s[i]} \leq E_{s},  \underset{j=1}{\overset{N}{\sum}} \frac{1}{T_u[j]} \leq E_{u}, \label{power_constraint}\\
~&T_{s}[i] \geq 0, \quad T_{u}[j] \geq 0.
\end{align}
Note that problem (P1.3) is a convex optimization problem, and thus can be optimally solved based on its Karush-Kuhn-Tucker (KKT) conditions. After some manipulations, the optimal solution can be obtained as
\begin{small}
 \begin{align}
T^*_{s,l}[i] &= \frac{ \underset{i=1}{\overset{N}{\sum}} \sqrt{\underset{j=i}{\overset{N}{\sum}} A_{s,l}[i][j]} }{ E_s \sqrt{\underset{j=i}{\overset{N}{\sum}} A_{s,l}[i][j]} }, \\ T^*_{u,l}[j] &= \frac{\underset{j=1}{\overset{N}{\sum}} \sqrt{\underset{i=1}{\overset{j}{\sum}} A_{u,l}[i][j]} }{E_u \sqrt{\underset{i=1}{\overset{j}{\sum}} A_{u,l}[i][j]} }, ~~i,j \in \cal N.
\end{align}
\end{small}

After solving problem (P1.3) with the given local points $T_{s,l}[i]$ and $T_{u,l}[j]$, the SCA algorithm proceeds by iteratively updating $T_{s,l+1}[i]=T^*_{s,l}[i], i \in \cal N$ and $T_{u,l+1}[j]=T^*_{u,l}[j], j \in \cal N$. According to  the SCA convergence result in \cite{beck2010sequential}, a monotonic convergence is guaranteed. The corresponding converged power allocation solutions can be retrieved via the equalities $P_s^*[i]=1/T_s^*[i]$ and $P_u^*[j]=1/T_u^*[j], i,j \in \cal N$.

\vspace{-0.3cm}
\subsection{UAV Trajectory Optimization}
For any given time-slot pairing and source/UAV transmit power, the sub-problem of UAV trajectory design is given by
\begin{align}
\text{(P1.4)}:\underset{\{x[i],y[i]\}}{\max}
~&\underset{i=1}{\overset{N}{\sum}} \underset{j=i}{\overset{N}{\sum}} ~w[i][j] f_3(x[i],x[j],y[i],y[j]) \nonumber\\
\mathrm{s.t.} ~~& (\ref{maxV1}), (\ref{maxV2}), (\ref{maxV3}). \nonumber
\end{align}
Here \begin{small}$f_3(\!x[i],x[j],y[i],y[j]\!)\!=\!\log_2(1\!+\!\frac{\beta_s[i]\beta_u[j]}{\beta_u[j]\theta_s[i]+\beta_s[i]\theta_u[j]\!+\!\theta_s[i]\theta_u[j]})$\end{small},
with \begin{small}$\theta_u[j]= (x[j]-L)^2+y^2[j]+H^2, \theta_s[i]= x^2[i]+y^2[i]+H^2, \beta_u[j]=P_u[j]\beta_0/\sigma^2$\end{small} and \begin{small}$\beta_s[i]=P_s[i]\beta_0/\sigma^2$\end{small}.

Problem (P1.4) is non-convex as the objective function is non-concave w.r.t. $\{x[i],y[i]\}^N_{i=1}$.
Nonetheless, similarly as the previous subsection, since $f_3(x[i],x[j],y[i],y[j])$ is jointly convex w.r.t. both $\theta_s[i]$ and $\theta_u[j]$, we can obtain a lower bound on $f_3(x[i],x[j],y[i],y[j])$ via its first-order Taylor approximation at the local point \begin{small}$(\theta_{s,l}[i],\theta_{u,l}[j])\!=\!(x_l^2[i]\!+\!y_l^2[i]\!+\!H^2,(x_l[j]-L)^2\!+\!y_l^2[j]\!+\!H^2)$\end{small} as
\begin{align}
&f_{3,\text{LB}}(x[i],x[j],y[i],y[j])\triangleq
f_3(x_l[i],x_l[j],y_l[i],y_l[j]) \nonumber\\
&-\!B_{s,l}[i][j](\theta_s[i]-\theta_{s,l}[i])\!-\!B_{u,l}[i][j](\theta_u[j]-\theta_{u,l}[j]),\label{traj.lb}
\end{align}
where
\begin{small}
\begin{align}
B_{s,l}[i][j]&=\frac{\beta_s[i]\beta_u[j]}{F[i][j](\beta_s[i]+\theta_{s,l}[i])\ln2 } > 0, \\ B_{u,l}[i][j]&=\frac{\beta_s[i]\beta_u[j]}{F[i][j](\beta_u[j]+\theta_{u,l}[j])\ln2 } > 0,
\end{align}
\end{small}
with\begin{small} $F[i][j] \triangleq \beta_u[j]\theta_{s,l}[i]+\beta_s[i]\theta_{u,l}[j]+\theta_{s,l}[i]\theta_{u,l}[j], i,j \in \cal N$\end{small}.

It is worth mentioning that the function $f_{3,\text{LB}}(x[i],x[j],y[i],y[j])$ given in (\ref{traj.lb}) is {\it not} the first-order Taylor approximation of the objective function $f_3(x[i],x[j],y[i],y[j])$ w.r.t. $(x[i],x[j],y[i],y[j])$. Nonetheless, it can be verified that this lower bound has the same objective value and gradient as $f_3(x[i],x[j],y[i],y[j])$ at the given local points $(x_l[i],x_l[j],y_l[i],y_l[j])$. As such, the SCA algorithm is still applicable to solve (P1.4) for local optimality\cite{beck2010sequential}.

Therefore, in the $l^{\text{th}}$ SCA iteration, we need to solve the following problem
\begin{align}
\text{(P1.5)}: \underset{\{x[i],y[i]\}}{\max}&\underset{i=1}{\overset{N}{\sum}} \underset{j=i}{\overset{N}{\sum}}
~w[i][j]f_{3,\text{LB}}(x[i],x[j],y[i],y[j])
\nonumber\\
\mathrm{s.t.}\;& (\ref{maxV1}), (\ref{maxV2}), (\ref{maxV3}). ~~~~~~~~~~~~~~~~~~~~~~~~\nonumber
\end{align}
Note that $f_{3,\text{LB}}(x[i],x[j],y[i],y[j])$ is a quadratic and concave function of $(x[i],x[j],y[i],y[j])$. As such, (P1.5) is a convex optimization problem, and thus can be optimally solved via the interior-point algorithm\cite{BoydBook}. Denote the optimal solution to (P1.5) as $\{x_l^*[i],y_l^*[i]\}$. The SCA algorithm proceeds by iteratively updating $x_{l+1}[i]=x_l^*[i]$ and $y_{l+1}[i]=y_l^*[i], i \in \cal N$ until the convergence condition is met.

\vspace{-0.3cm}
\subsection{Overall Algorithm}
Based on the above solutions, problem (P1) can be iteratively solved by applying the AO algorithm, i.e., optimizing the time-slot pairing, the source/UAV transmit power allocation and the UAV trajectory in an alternating  manner. The overall algorithm is summarized in Algorithm 1. It is easy to verify that the AO algorithm generates non-decreasing objective values of (P1), and thus is guaranteed to converge.
\begin{algorithm} [t]\label{alg:1}
\begin{small}
	\caption{AO Algorithm for Solving Problem (P1)}	
	\begin{algorithmic}[1]
       \State Set $l=0$. Initialize the source/UAV's power allocation $\{P_{s,0}[i],P_{u,0}[j]\}, i,j \in {\cal N}$ and the UAV's trajectory $\{x_0[i],y_0[i]\}, i\in \cal N$.
        \State \textbf{repeat}
		\State Solve (P1.1) via the Hungary algorithm and obtain the time-slot pairing solution as $w_{l+1}[i][j], i,j \in {\cal N}$.
		\State Solve (P1.2) via the SCA algorithm and obtain the souce/UAV transmit power solution as $\{P_{s,l+1}[i],P_{u,l+1}[j]\}, i,j \in \cal N$.
		\State Solve (P1.4) via the SCA algorithm and obtain the UAV's trajectory solution as $\{x_{l+1}[i],y_{l+1}[i]\}, i\in \cal N$.
       \State Update $l=l+1$.
		\State \textbf{until} the difference between the output from the current iteration and that from the last iteration is smaller than a prescribed constant $\epsilon$.
		\State Output $\{w[i][j],P_s[i],P_u[j]\}, i.j \in {\cal N}$ and $\{x[i],y[i]\}, i\in \cal N$.
	\end{algorithmic}
\end{small}
\end{algorithm}

\vspace{-0.3cm}
\section{Numerical Results}
In this section, numerical results are provided to evaluate the performance of our proposed SAF relaying protocol. Unless otherwise specified, the simulation settings are as follows. The distance between S and D is $L=2000$ m, i.e., S and D are located at $(0,0,0)$ and $(2000,0,0)$, respectively. The altitude of the UAV is fixed at $H=100$ m to ensure the LoS dominant links with S and D. The communication bandwidth per link is 20 MHz with the carrier frequency at 5 GHz. Without loss of generality, we assume the noise power spectrum density at UAV and D are equal and the value is -169 dBm/Hz. Thus, the reference SNR at $d_0=$ 1 m can be obtained as $\gamma_0=80$ dB. The UAV's maximum velocity is set to be $V_{\max}=40$ meters per second (m/s). The total flight period is $T=100$ s, and the length of each time-slot is 0.25 s, i.e., $N=400$ to ensure the accuracy of the time discretization. Accordingly, the maximum displacement of the UAV within a single time-slot is $D_u=10$ m. In the AO algorithm, we set $\epsilon=0.01$. The initial source/UAV power allocations are both assumed to be uniform over the $N$ time-slots, i.e., $P_{s,0}[i]=E_s/N, i \in \cal N$ and $P_{u,0}[j]=E_u/N, j \in \cal N$. The average transmit power of S and the UAV are assumed to be identical as $P=E_s/N=E_u/N$. The UAV trajectory is initialized as a straight line from S to D.

\begin{figure}[t]
\begin{center}
\subfigure[IAF]{
\includegraphics[width=1.6in]{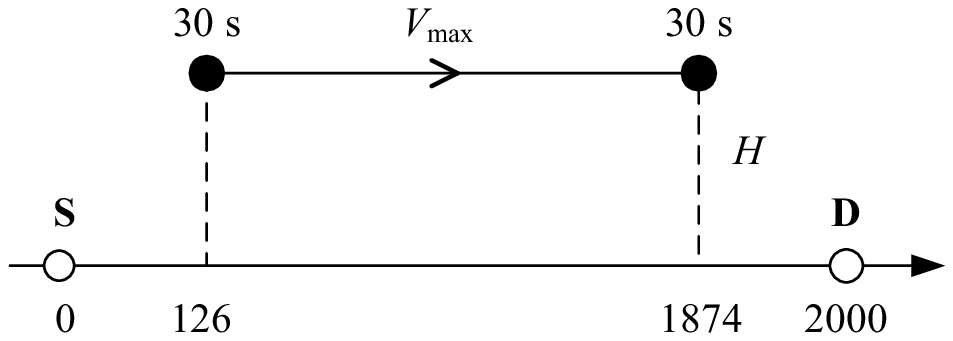}}
\subfigure[SAF / DF]{
\includegraphics[width=1.6in]{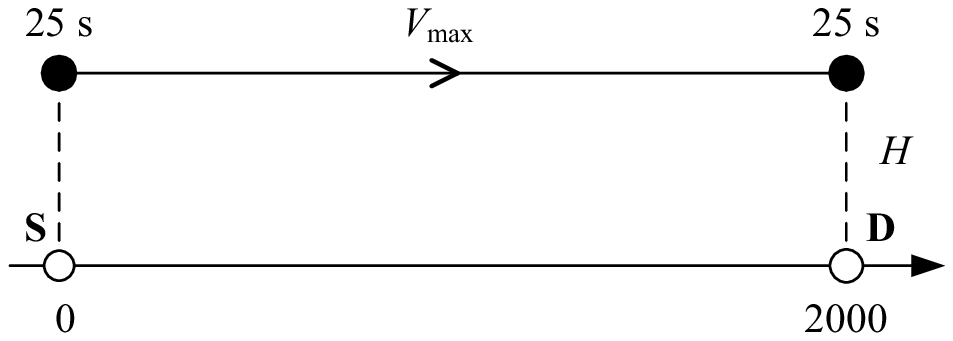}}
\caption{Comparison of optimized UAV trajectories under different relaying schemes.}\label{fig:trajectory}
\end{center} \vspace{-0.6cm}
\end{figure}

\begin{figure}[t]
\begin{center}
\includegraphics[width=3.2in]{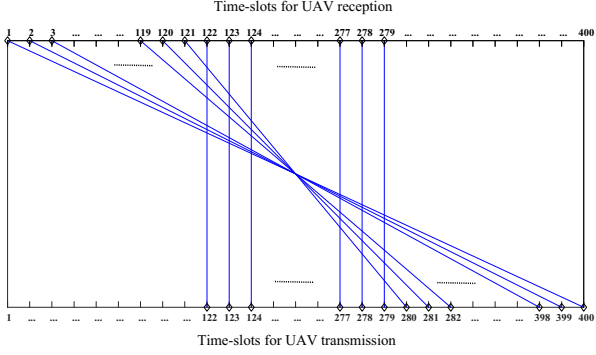}
\caption{Time-slot pairing for SAF relaying.} \label{fig:pairing}
\end{center}  \vspace{-0.6cm}
\end{figure}

\begin{figure}[t]
\begin{center}
\includegraphics[width=3.2in]{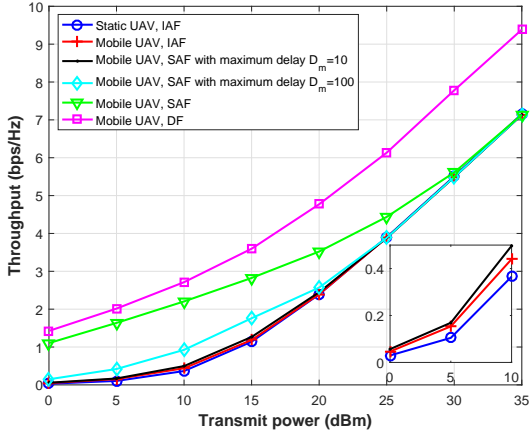}
\caption{Throughput versus UAV transmit power.} \label{fig:throughputwithpower}
\end{center}  \vspace{-0.6cm}
\end{figure}

First, to gain useful insights into the effect of UAV mobility, Fig.\,\ref{fig:trajectory} plots the optimized UAV trajectories under three different relaying schemes with free initial and final locations. The following schemes are considered, namely, the proposed SAF scheme, the conventional IAF scheme, as well as the DF scheme [2]. In this case, it can be easily proved that the optimal UAV trajectories must satisfy $y[n]=0$ and $0 \le x[n] \le L, \forall n \in \cal N$ under all considered schemes. The UAV's and source's transmit power are set to be identical as 15 dBm. From Fig.\, \ref{fig:trajectory}, it is observed that the UAV follows the ``hover-fly-hover'' trajectories in all the considered relaying schemes, i.e., it first hovers above an initial location, then flies towards a final location at its maximum speed, and finally hovers above the final location. Nonetheless, the hovering location and duration are observed to be different in general. Specifically, in the SAF and DF schemes, the UAV hovers above S and D for around 30 s to fully enjoy the best channel condition for signal reception and transmission, respectively. In contrast, in the IAF scheme, the UAV is observed to hover above two locations between S and D. This is expected as IAF requires the UAV to forward the received data packet instantly, hovering above S and D may compromise the quality of signal transmission to D and signal reception from S, respectively, due to the long distance between them.

Fig.\,\ref{fig:pairing} shows the time-slot pairing by the proposed SAF relaying scheme, with $P_s=P_u=15$ dBm. If two time-slots are paired, they are connected by a straight line (e.g., the $1^{\text{st}}$ and $400^{\text{th}}$ time-slots). Due to the size limit, we only show some of the paired time-slots. %
The average delay for all data packets is $30.25$ s with the proposed SAF under this setup.
It is observed that the UAV receives signal from S only in the first $279$ time-slots, while the UAV starts to forward the received signal to D from the $122^{\text{nd}}$ time-slot. Furthermore, the UAV forwards the signal received in the first 121 time-slots using the last 121 time-slots, while in the intermediate $122^{\text{nd}}$-to-$279^{\text{th}}$ time-slots, it forwards the signal received in the same time-slot, i.e., IAF is applied. This is because the UAV's channels to S and D are roughly equally good  during this period, as compared to the asymmetric channels with them in the other time-slots. The proposed SAF relaying scheme thus enables the UAV to schedule its signal reception and transmission more flexibly over its trajectory, as compared to the conventional IAF relaying.

Finally, in Fig.\,\ref{fig:throughputwithpower}, we plot the end-to-end throughput achieved by different relaying schemes.
In addition, we also plot the throughput of the proposed SAF scheme with the maximum allowable delay $D_m=10$ and $100$ (i.e., $2.5$ s and $25$ s), respectively (see Remark 1). From Fig.\,\ref{fig:throughputwithpower}, it is observed that the proposed SAF scheme outperforms the IAF relaying even with the given delay constraints, thanks to its flexible time pairing, as well as the AF static relaying by deploying the UAV at a fixed location (found via one-dimensional search). However, in the high transmit power regime, the performance gap between the proposed SAF and the conventional IAF diminishes, due to the more similar channel powers from the UAV to S and D regardless of its horizontal location.
It is also observed that as $D_m$ decreases, the throughput of SAF degrades, which implies an inherent throughput-delay tradeoff in SAF.
Last, it is observed that DF mobile relaying achieves the best performance as compared to all considered AF mobile relaying schemes, as the information decoding at the UAV relay mitigates the noise more efficiently than AF relaying.

\section{Conclusion}
This letter proposed a new SAF relaying protocol for UAV-enabled relaying systems by exploiting the UAV signal buffering for flexible transmit/receive  scheduling. The time-slot pairing, source/relay transmit power allocation and UAV trajectory were jointly optimized to maximize the end-to-end throughput. We proposed an iterative algorithm based on AO and SCA to obtain a high-quality suboptimal solution. Numerical results demonstrated that the proposed SAF relaying achieves significant throughput gains over the conventional IAF relaying, when the UAV/source transmit power or the system operating SNR is not high.

\end{document}